\documentclass[runningheads]{llncs}
\usepackage[T1]{fontenc}
\usepackage{amssymb}
\usepackage{graphicx}
\usepackage{hyperref}
\usepackage{listings}
\usepackage{xcolor}
\usepackage{xspace}
\usepackage{amsmath} 
\newcommand{\ANGLED}[1]{\ensuremath{\langle #1 \rangle}}

\newif\ifdraft

\drafttrue
\lstdefinelanguage{Fstar}{
  morekeywords={module, let, type, val, assume, begin, end, function, forall, fun, inline_for_extraction},
  sensitive=true,
  morecomment=[l]{//},
  morecomment=[s]{(*}{*)},
  morestring=[b]",
}

\begin{document}
\title{Enforcing MAVLink Safety \& Security Properties Via Refined Multiparty Session Types}
\titlerunning{Enforcing MAVLink Safety Properties via RMPSTs}

%\orcidID{0009-0002-0284-4787}
\author{ 
Arthur Amorim\inst{1}\orcidID{0009-0003-7712-5055} \and
Max Taylor\inst{2} \orcidID{0009-0005-7873-9694} \and \\
Trevor Kann \inst{3}\orcidID{0009-0004-5197-2448} \and
William L. Harrison\inst{2}\orcidID{0000-0002-3760-3556} \and \\
Gary T. Leavens\inst{1}\orcidID{0000-0003-3271-3921} \and
Lance Joneckis\inst{2}\orcidID{0009-0002-0284-4787}
}

\authorrunning{A. Amorim et al.}


\institute{ 
University of Central Florida, Florida, USA\\
\email{\{arthur.amorim, leavens\}@ucf.edu}
\and
Idaho National Laboratory, Idaho Falls, USA\\
\email{\{maxwell.taylor,william.harrison,lance.joneckis\}@inl.gov}
\and 
Carnegie Melon University, Pennsylvania, USA \\
\email{tkann@cmu.edu}
}

\maketitle              % typeset the header of the contribution
\begin{abstract}
%The abstract should briefly summarize the contents of the paper in
% 1. State the problem
A compromised system component can issue message sequences that are  legal while also leading the overall system into unsafe states.
% 2. Say why it's interesting
Such stealthy attacks are challenging to characterize, because message interfaces in standard languages specify each individual message separately but do not specify safe sequences of messages.
%3. Say what your solution achieves
We present initial results from ongoing work applying refined multiparty session types as a mechanism for expressing and enforcing proper message usage to exclude unsafe sequences.
% 4. Say what follows from your solution 
We illustrate our approach by using refined multiparty session types to mitigate safety and security issues in the MAVLink protocol commonly used in UAVs.
\end{abstract}

\section{Introduction}

%\maxt{\emph{What are you trying to do?}}
Attackers who are already inside a system can mount stealthy attacks in which they send apparently innocuous messages that may still cause damage. 
We aim to limit their damage by dynamically checking that messages conform to a protocol,
enabling critical systems to have verified safety guarantees.

%\maxt{\emph{How is it done today? Limits of current practice}}
One approach to keeping attackers out of critical systems is exemplified by the DARPA High-Assurance Cyber Military Systems (HACMS) program~\cite{HACMS}. 
HACMS combines SMACCMPilot~\cite{smaccmpilot} and seL4~\cite{sel4} to prevent external adversaries from compromising the autopilot system of a UAV.  

However, researchers have since identified novel attacks against UAVs~\cite{RVFuzzer,kim2021pgfuzz}. % [[[How is this different than the attacks that prompted the  project? -- Trevor ]]]
%\edit{Is ``stealthy attack'' a term of art, or is it something you are defining?} % It is a term of art. This is an example of a stealthy attack. I removed the italics so that it doesn't seem like we're defining a stealthy attack this way.
The main such attack is a \emph{stealthy attack}, in which a compromised ground control station (GCS) sends seemingly innocuous messages to the UAV. 
The GCS might, for example, be following the dictates of a disgruntled employee. % [[[Really akward phrasing -- Trevohttps://www.overleaf.com/project/67350802dbc051777dd75063#r]]]
Such messages take advantage of an incorrectly designed or implemented protocol to evade intrusion detection and still cause unsafe behavior.  %[[[ I don't think you want to focus on intrusion detection here for a couple reasons: one, you haven't really talked about it before, two your system is more general than this, and three it's kind of out of scope/orhtogonal to this work -- Trevor]]]
As a response to stealthy attacks, researchers have proposed runtime verification~\cite{protoARMA,uasVerification,icss24}.

%\maxt{\emph{What is new in your approach?}}
Our approach, which we call Dynamically Assured Typed Universal Messaging (DATUM),  checks that the communication between the controller and the controlled system (e.g., the GCS and the UAV) conforms to a specified protocol.
The protocol itself can also be checked to validate that it preserves safety.\footnote{We do not discuss the details of checking the safety of protocols in this paper.}

In this paper, we formalize a substantial portion of the MAVLink protocol~\cite{mavlink} used by GCSs to control UAVs. 
MAVLink specifies more than 200 messages and 100 custom data types. 
% [[[Is the following sentence necessary or useful? -- Gary]]]
From the MAVLink specification, we generated more than 8k lines of code. 

%\maxt{\emph{Who cares?}}
Our approach benefits two audiences.
First, operators can use our approach to mitigate the risks posed by stealthy attacks while they await patches.
Second, engineers can validate a patch's modifications  before releasing it.
%\trevor{I like this paragraph (maybe switch the order?)}

%We validate the effectiveness of our approach using three case studies taken from real-world safety \& security issues.

%\maxt{Contributions.}
To summarize our contributions, we:
\begin{itemize}
    \item Describe a methodology that can prevent stealthy attacks against UAVs.    %\trevor{Is it not the case that your contribution is actually stronger than this and this is only the motivation?}
    % I think it is stronger, but we don't elaborate on those strengths in this paper. So, I hesitate to claim those contributions.
    \item Present three case studies showing how our formalization mitigates real-world safety and security issues in MAVLink. We evaluated our case studies using two widely deployed autopilots, ArduPilot and PX4~\cite{ardupilot,px4}.
    \item  
    Formalize a substantial part of the MAVLink protocol~\cite{mavlink} using DATUM. 
\end{itemize}

\section{Background}
\subsection{Session Types}
DATUM uses session types, a rich typing discipline~\cite{honda1993}. 
Binary session types can enforce constraints, such as message ordering, and capture temporal properties between two actors, such as the absence of deadlocks~\cite{bst}.
Multiparty session types (MPSTs) extend these properties to systems with multiple actors~\cite{mpsts,hondaMultiPartyAsync}. 
For each exchanged message, MPSTs can express the message sender, receiver, and payload type.
MPSTs have led to many real-world applications~\cite{spy,Bocchi2017,Scribble}. 

However, MPSTs cannot reason about payload values, as they cannot constrain the values being exchanged in the systems. 
For example, MPSTs cannot write a protocol where $A$ sends $B$ a number $n$, and then $B$ sends that same $n$ to $C$.
Refined multiparty session types (RMPSTs)~\cite{2024_refinements,rmpsts,rmpsts_phd} can specify such constraints by expressing properties of values within messages. 
While RMPSTs are often used in static reasoning; we are the first to use them for dynamic checking.

An RMPST has a syntax that follows one of the cases given below:
% [[[A grammar would be more precise. Perhaps give a reference to a paper with the grammar? -- Gary]]]
\begin{itemize}
  \item \textbf{Messages:} A message specifies a sender (e.g., $A$), a receiver (e.g., $B$), a label (e.g., \textit{MSG}), and the message's payload (e.g., $x$). The payload is specified with a type (e.g., $\mathbb{N}$) and may optionally have a refinement, (e.g., $x > 7$). %The following is an example:
  {\small
  \[
    A \to B: \textit{MSG}(x: \mathbb{N} \{x > 7\})
  \]
   }

  \item \textbf{Guarded Choice:} A refinement may contain a choice that depends on values. The following is an example where $B$ sends to $A$ different messages depending on the refinement: {\small
  \[
    B \to A:
    \begin{cases}
      \texttt{OPTION1}(y: \mathbb{N} \{x = 10\}) \\
      \texttt{OPTION2}(z: \mathbb{N} \{x \neq 10\})
    \end{cases}
  \]
  }
  % [[[Should the syntax have a matching right bracket? I seem ambiguous otherwise. -- Gary]]]

  \item \textbf{Mu-Recursion:} A recursive protocol can be specified using the $\mu$ operator. For example, the following recursive protocol has $n < 5$ as its recursion condition, where $n$ is incremented after every recursion.
{\small
    \[
    \begin{aligned}
    \mu.\texttt{T}&(n: \mathbb{N} \{0 \leq n \land n < 5\} ) \ANGLED{n = 0} \\
    &A \to B:\textit{MSG}(y: \textit{bool} \{y = true\}).\texttt{T} \ANGLED{n = n + 1}      
    \end{aligned}
    \]
  }

  \item \textbf{End:} The \textbf{End} construct specifies a protocol's endpoint. The overall protocol terminates when all participants reach their endpoints. For example: {\small
  \[
    A \to B: \texttt{FINAL\_MSG}(t: \texttt{STATUS} \{t = \texttt{SUCCESS} \lor t = \texttt{FAIL}\}). \textbf{End}
  \]
  }
\end{itemize}

\subsection{The F* Interactive Theorem Prover}
F* is an interactive theorem prover that integrates the Z3\cite{z3} SMT solver for automated proofs, while also allowing users to write constructive proofs\cite{fstar}.
This combination allows trivial properties to be proved with minimal input, while allowing for interactive proofs of other properties. 
Both kinds of properties can be proven to hold in all possible executions. 

We chose to represent our RMPSTs in F* for its proof automation and native support for refinements.
% [[[Section 3.3 says that F*'s native refinements are NOT used, so we shouldn't say that F* was chosen partly for using such refinements.]]]
% [[[We use a subset of those refinements. So thye are still used but not unrestrictedly.]]]
Previous work showed promise in the formalization of a binary-session-typed calculus in theorem provers~\cite{pi-calc_CoQ,QTT-sess-Lean}. 
Other works showed how to prove properties about session subtyping~\cite{Completeness_Coq}. 
Zooid~\cite{Zooid} is a domain specific language embedded in  Coq~\cite{coquand1986calculus} that allows users to define protocols using MPSTs.
Zhou et. al. showed how an F* API can be used to statically verify an RMPST protocol specification~\cite{rmpsts}.
DATUM allows for protocols to be described as F* terms, allowing runtime traces to be checked.

\section{Methodology}

\subsection{Case Study I: Resource Usage}

We demonstrate our methodology for preventing stealthy attacks against the MAVLink mission sub-protocol.
MAVLink is the \textit{de facto} standard for GCS-UAV communications despite numerous safety and security shortcomings~\cite{allouch2019mavsec,hamza2024mavlink,kwon2018empirical,taylor2021study}. 

MAVLink allows the GCS to send mission directives to the UAV. Mission directives include waypoints, geofences, and safepoints. 
% Correct implementation of mission directives is vital to the UAV's safety. 
The GCS starts the mission sub-protocol by sending the \lstinline{MISSION_COUNT} $N$ MAVLink message, where $N$ is the number of mission items the GCS will send. 
Consequently, the UAV expects exactly $N$ mission items.  

%Thus, it should reject mission items when it receives a number of mission items different than $N$. Unfortunately, rejecting mission items is not the standard mission sub-protocol behavior. 
Issue reports show how a UAV running ArduPilot accepted fewer mission items than its declared mission count~\cite{MissionPlannerIssue1248}. 
This error was caused by a failed waypoint upload from an earlier flight. 
These uploads were kept in a buffer until the next use of the mission sub-protocol. 
Midway through the mission, the UAV began following a previous mission's flight plan. 
Such unexpected changes in the flight plan could have compromised the safety of the UAV.
The mission sub-protocol should have the capability to reject missions based on missing items. 

\subsection{Mitigation With RMPSTs}

% RMPSTs allow for the implementation of recursive protocols, which are specified with the $\mu$-recursion operator. 
% This operator is especially useful for recursive protocols with break conditions, such as the \textbf{End} in the RMPST below.

In our methodology, the sub-protocol initiated by the message \mbox{\lstinline{MISSION_COUNT} $N$} is specified as an RMPST (shown below) with recursion over $curr : \mathbb{N}$ guarded by the refinement $\{0 \leq curr \land curr \leq N\}$. 
% [[[I think the following sentence is unnecessarily redundant. -- Gary]]]
%Refinement types like this are a  type and a predicate that restrict the values that inhabit the type. 
The recursive protocol offers a choice between the messages \lstinline{MISSION_REQUEST_INT} and \lstinline{MISSION_ACK}. 
The \lstinline{MISSION_}\lstinline{REQUEST_INT} message must have a payload between $0$ and $N-1$ and the GCS must request missions sequentially. 
We enforce the second condition by comparing the payload $x$ to $curr$, which increments after every iteration. 
Following the \lstinline{MISSION_REQUEST_INT} message, the UAV sends a \lstinline{MISSION_ITEM_INT} message and recurs while incrementing the $curr$ variable. 

The sub-protocol ends with a \lstinline{MISSION_ACK} message and a \lstinline{MISSION_RESULT} payload $t$.
When an error occurs, the UAV sends a \lstinline{MISSION_ACK} message with the respective \texttt{ERROR} payload.
If all missions upload successfully, $curr$ reaches value $N$ and the \lstinline{MISSION_ACK} message represents a correct sub-protocol run.
{\small
\[
\begin{aligned}
   &\textbf{GCS} \to \textbf{UAV:}\ \texttt{MISSION\_COUNT}(N: \mathbb{Z}\ \{N \geq 1 \land N < \texttt{MISSION\_ITEM\_LIMIT}\})\\
   &\textbf{$\mu.$T}(curr: \mathbb{Z} \{ 0 \leq curr \land curr \leq N\}) \ANGLED{curr = 0} \\
   & \quad\textbf{UAV $\to$ GCS:}
   \begin{cases}
      \texttt{MISSION\_REQUEST\_INT}(x: \mathbb{Z} \{curr< N \land x=curr\})\\
      \texttt{MISSION\_ACK}(t: \texttt{MISSION\_RESULT} \{t = \texttt{ERROR}  \lor curr = N\}). \textbf{End} 
    \end{cases} \\
    & \quad\textbf{GCS $\to$ UAV:}\ \texttt{MISSION\_ITEM\_INT}(y: \mathbb{Z} \{y=x\}).\textbf{T} \ANGLED{curr=curr+1}
\end{aligned}
\]
}
\subsection{Using DATUM To Write RMPSTs in F*}
\label{sec:: DATUM RMPST}

\autoref{fig:case-study-iii-formal} shows the RMPST from our case study formalized in F*. This listing relies on the DATUM framework that we have developed. The helper type \lstinline{mission_req} shows how we can represent the refinement on the \lstinline{MISSION_}\lstinline{REQUEST_INT} GCS message that appears in the case study. Other helper types like \lstinline{mission_item} are similar. \autoref{fig:case-study-iii-formal} excludes their definitions for brevity.

\begin{figure}[t]
    \centering
    \begin{lstlisting}[language=Fstar,basicstyle=\tiny\ttfamily]
type mission_req (curr : int) (n : int) = hidden (stamped_val #mavlink_user_type
 (RefType #_ #StampUserType 
  ($\lambda$ (v : R.prim_val M.mavlink_user_type R.StampUserType) $\rightarrow$ UserType? v && 
    MissionRequestInt? (UserType?.v v) && (MissionRequestInt?.v (UserType?.v v)).seq = curr && curr < n)))
let mission_proto = 
 Offer gcs uav [
  choice "MISSION_COUNT" ($\lambda$ (msg : mission_count_msg) $\rightarrow$
    Mu (iteration_limit (get_count msg)) (Val (Int 0))
        ($\lambda$ curr $\rightarrow$ Offer uav gcs [
          choice "MISSION_REQUEST_INT" (
           $\lambda$ (x : mission_req (Int?.v (Val?.v curr)) (get_count msg)) $\rightarrow$ 
              Offer gcs uav [
               choice "MISSION_ITEM_INT" (
                $\lambda$ (_ : mission_item (Int?.v (Val?.v curr))) $\rightarrow$
                 Recur 0 (Val (Int (Int?.v (Val?.v curr) + 1))))]);
          choice "MISSION_ACK" ($\lambda$ (_ : mission_ack (Int?.v (Val?.v curr)) (get_count msg)) $\rightarrow$ End)]))]
    \end{lstlisting}
    \caption{The RMPST from our case study, written in F* using DATUM.}
    \label{fig:case-study-iii-formal}
\end{figure}

Notably, DATUM does not utilize F*'s native refinement types when writing an RMPST. Utilizing F*'s native refinement types would require expensive dynamic type-checking of the entire F* programming language to attest that a communication conforms to an RMPST. Instead, DATUM provides the \lstinline{RefType} constructor, shown in \autoref{fig:case-study-iii-formal}. This reduces the dynamic attestation problem to checking that a predicate evaluates to true.

DATUM provides sound support for modeling RMPSTs. In DATUM's model, RMPSTs have four constructors: \lstinline{Offer}, \lstinline{Mu}, \lstinline{Recur}, and \lstinline{End}. Both \lstinline{Offer} and \lstinline{Mu} introduce new variables. 
New variables are represented as function parameters.

DATUM uses F*'s effect system to prevent function parameters representing RMPST variables from creating ill-formed RMPSTs. 
F*'s effect system is an extensible lattice of monadic effects \cite{fstar}.  
DATUM defines a \lstinline{hidden} monad that prevents variables from being used outside of refinements. A value of type \lstinline{hidden $\alpha$} cannot be used in F*'s default effect \lstinline{Tot}. However, DATUM can use it  when performing I/O during dynamic attestation.

\begin{figure}[b]
    \centering
    \includegraphics[width=\linewidth]{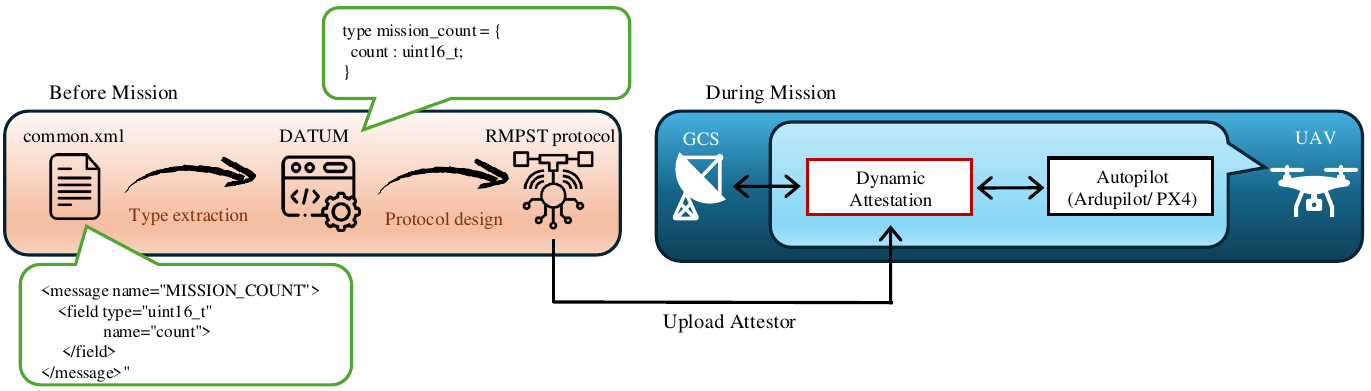}
    \caption{Engineers use RMPSTs from the DATUM MAVLink specification to create a protocol that mitigates safety and security issues. At runtime, DATUM checks on the UAV that messages conform to the protocol.}
    \label{fig:overview}
\end{figure}

DATUM automatically generates a dynamic attestation tool from the specification of MAVLink, as shown in Figure \ref{fig:overview}. When using DATUM, communication between the GCS and the autopilot software passes through a MAVLink proxy. The MAVLink proxy uses the attestation tool to check each message. Non-compliant messages identified by DATUM are reported to the UAV's pilot.

\subsection{Inside of DATUM}

\autoref{fig:overview} shows an overview of our approach. First, we automatically generate F* types from MAVLink's machine-readable specification file. Then, we use DATUM to write an RMPST describing MAVLink. Inside the MAVLink RMPST, we include refinements that mitigate vulnerabilities.

The MAVLink specification~\cite{mavlink} defines more than 300 types to describe messages and their payloads. These types must be written in F* to use DATUM to define RMPSTs. Manually writing these 300 types in F* would be tedious. Instead, we use automation to generate F* types. Consider the \lstinline{MISSION_COUNT} message, whose F* and DATUM representation is shown in the left call-out of \autoref{fig:overview}. This code is automatically generated from the \lstinline{common.xml} file that MAVLink distributes. The right call-out in \autoref{fig:overview} shows the part of \lstinline{common.xml} that contributes to the generation of the \lstinline{mission_count} type. In total, automation generates $\approx 8,500$ lines of F* code.

% We automatically generate F* bindings for MAVLink. The MAVLink protocol maintainers distribute a file called \lstinline{common.xml}. \lstinline{common.xml} is a machine-readable file containing descriptions of MAVLink's data types and messages. We use the data type descriptions to generate corresponding F* data types. We use the message descriptions to generate F* record types corresponding to each MAVLink message.

\subsection{Additional Case Studies}

In this section we discuss two additional case studies and demonstrate their respective mitigation using RMPSTs. For brevity, we omit their formalization.
%This design can help mitigate the effects of input-validation, policy violation, and sub-protocol bugs.

%Some examples involve protocols used by flight control software, such as PX4, while others involve bugs in the MAVLink protocol %[[[Give references for MAVLink?]]]
%that is used for lower-level communication between a GCS and a UAV. 

\subsubsection{Case Study II: Overly Permissive Protocol}
An overly permissive protocol bug occurs when a protocol allows values that may exceed safe ranges. 
The PX4 autopilot software includes over 200 configurable parameters. 
These parameters are used by PX4 to navigate the UAV. %Parameters change according to messages issued by the GCS in response to flight conditions.  However, 
Some parameters have interdependencies;
that is, the safe range of a parameter may depend on the values of related parameters. For example, in PX4 the \lstinline{MC_PITCHRATE_FF} and \lstinline{MC_PITCH_P} parameters represent the feed-forwarding pitch rate and pitch gain. 
When both parameters are set to high values, the UAV makes jerky movements to try to stabilize itself, 
but such movements can result in loss of stability. 
In these situations, a low maximum pitch rate (\lstinline{MC_PITCHRATE_MAX}) value is needed to ensure stability.

%Although the valid upper bound for this parameter is set to 1800 deg/s, values as low as 221 deg/s can cause the UAV to lose control~\cite{RVFuzzer}. 

%%This is an example of an \emph{overly permissive protocol} bug.

Overly permissive protocol bugs can lead to scenarios where the UAV crashes due to communications that do not violate the parameter's range requirements. 
This vulnerability could be used by a stealthy attacker. 
Such an attacker could send malicious messages to the UAV to change these parameters with the message \lstinline{PARAM_SET}, while still following the MAVLink protocol~\cite{RVFuzzer}. %An attacker could thus compromise the safety of the UAV while still following the protocol.
%Decreasing the upper bound for \lstinline{MC_PITCHRATE_MAX} is not a proper mitigation, because it would affect the UAV's performance under safe operations. 
However, RMPSTs can constrain dynamic parameter changes with respect to value dependencies. 
For example, this value dependency can be expressed  as follows (where $p$ and $q$ are weights for the parameters involved):

{\small
\[
\begin{aligned}
   \textbf{GCS $\to$ UAV:}\ &\texttt{PARAM\_SET(MC\_PITCHRATE\_MAX}, n: \mathbb{N} \\ 
   &  \{n < (p \times \texttt{MC\_PITCH\_P})\ \times (q \times \texttt{MC\_PITCHRATE\_FF})  \})
\end{aligned}
\]
}

The dynamic attestation checker applies the value dependency predicate in the refinement to the message payload $n$. 
Thus DATUM enforces that %\lstinline{MC_PITCHRATE_MAX} upper bound is inversely proportionate to the current values of \lstinline{MC_PITCH_P} and \lstinline{MC_PITCHRATE_FF}. 
%These parameters are tracked by the dynamic attestation checker as part of the UAV's state.
%Thus, 
the \lstinline{PARAM_SET} message is only seen by the UAV if it 
is safe to execute.

\subsubsection{Case Study III: Precondition Violations}
Some MAVLink messages, when executed at the wrong time, can cause harm to the UAV, however message preconditions are not enforced.
One such message, \lstinline{MAV_CMD_DO_PARACHUTE}, deploys the parachute in ArduPilot. 
The documentation defines the necessary preconditions for this message: the UAV is not ascending, the payload $n$ equals 2, its motors are armed, it is not in ACRO or FLIP mode, and its altitude is above \lstinline{CHUTE_ALT_MIN} parameter.
Deploying the parachute while violating any of the preconditions, even though they are not enforced, can lead to a crash. 
Thus, an attacker could wait for an opportunistic moment to deploy the parachute~\cite{kim2021pgfuzz}. 
%The violation of any preconditions could lead to an unsafe situation. 
%Even a seemingly innocuous message could cause the UAV to crash. 
%A crash was shown to happen in a study which describes protocol polices with temporal logic~\cite{kim2021pgfuzz}. 
%That study identified the bug, but offered no practical solution for preventing it.

%A model's state is an assignment of values to variables. RMPST is able to constrain messages based on refinements on values. Thus, RMPST can impose constraints on messages that reason about the UAS's state. 
DATUM can enforce these precondtions using RMPSTs, as shown below.
{\small
\[
\begin{aligned}
   &\textbf{GCS $\to$ UAV:}\ \texttt{MAV\_CMD\_DO\_PARACHUTE}(n: \mathbb{N} \{n=2\ \land motors = armed\ \land\\
   &mode \neq \texttt{ACRO} \land  mode \neq \texttt{FLIP} \land v\_z < 0 \land alt <= \texttt{CHUTE\_ALT\_MIN}\})
\end{aligned}
\]
}
%The current mode and armed status are obtained from the heartbeat and arm protocols. The \proxy can find out if the UAV is climbing by monitoring its z-velocity. Lastly, \lstinline{CHUTE_ALT_MIN} is a global variable. Therefore, a \lstinline{MAV_CMD_DO_PARACHUTE} message with payload ``2'' for parachute release can only be sent if it satisfies its refinement.  Consequently, \proxy will prevent such time-dependent precondition bugs.
\section{Evaluation}

We evaluated our methodology by retrofitting systems equipped with ArduPilot 4.5.7~\cite{ardupilot} and PX4 1.15.2~\cite{px4}. ArduPilot and PX4 were evaluated using their software in the loop (SITL) systems. All experiments were conducted on a machine equipped with an Intel i9 processor, 64 GB of RAM, and Ubuntu 24.04.

In our evaluation, we simulated a mission with 100 waypoints, each uploaded by the GCS to the autopilot software. 
Our experiments measured memory usage and message latency.\footnote{
To measure message latency, we recorded the timestamps when each message was sent and received. 
This required modifying the flight control software to record the appropriate timestamps.} 
The results of our experiments are shown in \autoref{tab:quant_results}. 

\begin{table}[t]
    \centering
    \begin{tabular}{|c|c|c|c|c|}
        \hline
         \textbf{Software} & \textbf{RSS} & \textbf{RSS w/ Datum} &\textbf{Latency} & \textbf{Latency w/ Datum}  \\
         \hline
         ArduPilot & 8,096 KB & 21,824 KB & 29.60 $\pm$ 6.09 $\mu s$ & 56.74 $\pm$ 6.75 $\mu s$ \\
         PX4 & 18,490 KB & 32,208 KB & 20.96 $\pm$ 1.97 $\mu s$ & 57.70 $\pm$ 6.53 $\mu s$ \\
         \hline
    \end{tabular}
    \caption{We measured memory and time by looking at the resident set size (RSS) (i.e., the physical memory the process is using) and message latency respectively.}
    \label{tab:quant_results}
\end{table}

The overhead numbers are  small when read as absolute figures. 
However, relative to the values without dynamic attestation the overhead seems high. 
We note that the relative overhead would be smaller in an actual deployment where the communication latency would be much greater than in our evaluation.  
(Our evaluation setup consisted of a single machine with communications exchanged on its loopback interface. Thus, communication latency was small, less than $30 \mu s$; however, latency between the GCS and UAV is typically 2-5 ms.)
\section{Conclusion}
In this work, we have shown how to use lightweight formal methods to retrofit existing systems to mitigate real-world safety and security weaknesses in a widely deployed protocol. 
Our methodology, DATUM, uses RMPSTs to write protocols with logical constraints. 
These constraints are dynamically checked. 
DATUM allows GCS software to prevent attacks that use unpatched bugs and to fine-tune protocols for future patches.

\bibliographystyle{splncs04}
\bibliography{bib}

\end{document}